\documentclass{article}
\usepackage[utf8]{inputenc}
\usepackage{latexsym,float}
\usepackage{authblk}
\usepackage{epsfig}
\usepackage{epstopdf}
\usepackage{amsmath,hyperref}
\usepackage{amssymb}
\usepackage{graphicx}
\usepackage{eepic}
\usepackage{bbm}
\usepackage{textcomp}
\usepackage{xcolor}
\usepackage[T1]{fontenc}

\title{Geodesically complete cyclic cosmologies and entropy}

\begin{document}
\author[1]{Petar Pavlović \thanks{petar.pavlovic@kozmologija.org}}
\author[2,1]{Marko Sossich \thanks{marko.sossich@fer.hr}}
\affil[1]{\textit{\small{Institute for Cosmology and Philosophy of Nature, Trg svetog Florijana 16, Križevci, Croatia}}}
\affil[2]{\textit{\small{University of Zagreb, Faculty of Electrical Engineering and Computing, Department of Physics,
	Unska 3, 10 000 Zagreb, Croatia}}}

\maketitle

\begin{abstract}
We demonstrate that there exists a class of cyclic cosmological models, such that 
these models can in principle solve the problem of the entropy growth, and are at the same time geodesically complete. We thus show that some recently stated conclusions, according
to which cyclic cosmologies solving the problem of entropy growth can not be 
geodesically complete due to the Borde-Guth-Vilenkin (BGV) theorem, are not justified. 
{We also add a short conceptual discussion on entropy and cyclic cosmology, and present a detailed analysis of entropy density growth during periodic and non-periodic evolution for cyclic cosmologies.}  

\end{abstract}

\section{Entropy and cyclic cosmology}
Cyclic cosmology represents an alternative to the big bang cosmology, replacing the
initial singularity with the infinite number of transitions from contraction to 
expansion of the Universe. As a theory, cyclic cosmology has many strong 
philosophical, logical and physical advantages. Cyclic cosmology avoids the problem 
of creation \textit{ex nihilo}, describes the Universe as infinite in time, gives 
natural and continues evolution of the Universe without singularities, while also 
solving the additional problems of the standard cosmological model, such as the 
horizon problem. For more details regarding the cyclic cosmology in the context of 
the open problems of modern cosmology see Introduction in \cite{nas1} and \cite{nas2}.
Starting from the early proposal of Tolman \cite{tolman}, there were many proposed 
models of cyclic universe, based on different specific assumptions and theoretical 
additions to general theory of relativity and standard cosmological model, specially 
in the last decades \cite{jen, dva, tri, cetiri, pet, sest, sedam, osam}. Recently, 
a general model-independent approach to study cyclic cosmologies was proposed
in \cite{nas2} and \cite{nas1}, and also applied to the problem of magnetogenesis 
\cite{natty}, cosmological constant problem \cite{nas3} and creation of wormholes 
around the cosmological bounce \cite{nas4}. \\ \\
Ever since Tolman started to investigate the possibility of cyclic cosmology 
\cite{tolman}, a so called entropy problem of the cyclic cosmology was pointed out. 
Namely, as the entropy of the Universe needs to increase both during the contracting 
and expanding phase, one would expect that it needs to grow without any limit in the 
Universe infinite in time, thus subsequently leading to its thermal death. This 
problem was even more severe in the original Tolman's proposal, since it was based on 
a spatially finite geometry of the Universe ($k=1$), where the radius of the Universe
is proportional to its entropy  -- and would then need to be practically singular in 
the past. Apart from this problem of the entropy growth, the contracting phase of the
cyclic cosmologies could be connected with additional problems of growing 
instabilities and black hole formation {\cite{lif, mix, dicke}}. In this work, we will not study these 
additional potential problems in detail, leaving this for future research, but we 
will focus only on the principal question of the entropy growth. \\ \\
Before proceeding further we should stress that invoking the concept of the entropy 
growth to question the possibility of cyclic cosmology is highly 
questionable. This follows from the fact that we at this moment do not have any proper
knowledge of how meaningful is the concept of the entropy growth when applied to the
whole Universe and its evolution, and when, furthermore, considered in a general 
framework of curved spacetime. It should also not be forgotten that the concept of entropy growth has a completely different character from the usual physical laws, since it is purely statistical in its essence, and thus represents a highly likely tendency, but not necessity -- which is furthermore dependent on the physical context of available phase space and its evolution. For this reason, it is highly problematic to skip the serious and detailed analysis of equations of motion and their solutions, and to make conclusions about the physics of the Universe using the simple shortcut -- "entropy needs to grow", as it is unfortunately sometimes done. There are further strong reasons to conclude that the usual concept of entropy and its growth could be problematic in the context of cosmology, or at least that we currently do not have a proper understanding of it. Namely, the evolution of the Universe does not seem to be in accord with the notion of the necessary entropy growth even in the standard cosmological model. The CMB spectrum, almost perfectly corresponding to the spectrum of the black body radiation, tells us that the early Universe was already in the state of thermodynamic equilibrium, which means almost maximal entropy. This is in a direct contradiction with the assumption that the early Universe was in the state of low entropy which is increasing \cite{calender, wald}. The proposals which invoke the concept of entropy of gravity to solve this issue \cite{penrose} are still very hypothetical and far from obvious. To conclude, it should be completely obvious that arguments based on the entropy growth can not be used to dismiss the viability of cyclic cosmology, since at the same time similar considerations lead to contradictions in the standard big-bang cosmology. {In fact, some interesting works and models regarding entropy considerations in cyclic cosmology can be found in \cite{barrow1, clifton, barrow2, ganguly}}.\\ \\
Having made this critical remarks, which show that entropy considerations are in fact not a justified argument against cyclic cosmology on the fundamental level, in this work we will,  for the sake of an argument, assume that the concept of entropy and its growth can be meaningfully applied to cosmology in a simple and usual manner. In this context, it was recently proposed that a cyclic cosmology can be made consistent with the entropy growth, invoking an exponential growth of the scale factor from one cycle to the next one \cite{steinhardt1, steinhardt2}. In this proposal, the total entropy of the Universe continues to grow between the cycles, but its density decreases due do the huge increase of the scale factor. Since the large regions of the Universe exit the causal horizon during such expansion, the entropy of some causally connected region in such a cyclic Universe can then be much smaller at the beginning of a new cosmological cycle. \\ \\
However, it was pointed out that such cyclic cosmology can not be geodesically complete due to the BGV theorem \cite{bgv1,bgv2,bgv3}. This theorem, proposed by Borde, Guth and Vilenkin, states that a sufficient condition for geodesic incompleteness is the positive value of average Hubble parameter.
It can thus be shown \cite{bgv2,bgv3} that the model proposed in \cite{steinhardt1} is past geodesically incomplete, since the decrease of entropy density in this model leads to the positive average value of the Hubble parameter. {One can {find} some examples 
where authors have simply added a constant to the scale factor which could potentially
circumvent the assumptions of the BGV theorem. However, those models are limited to the fact that the spatial curvature $k$ is nonzero. Due to this fact, the Friedmann equations could
admit a non-homogeneus solution of the differential equation, and thus a 
solution with a constant can be simply constructed mathematically \cite{osci1,osci2,osci3}.
Those results were, however, quantum-mechanically unstable and 
can collapse by quantum tunneling to zero radius as shown in \cite{osci4}.
From a more technical perspective, quite recently, the authors in \cite{easson} proved that an uncountable infinite class of classical solutions (monsters) which have $H_{av} \geq 0$ are in fact geodesically complete, while also those solutions rely on a more general definition of $H_{av}$ {\cite{kinn}.}
In our work we will also give a more general description of the cyclic Universe and show that some claims made in \cite{bgv2,bgv3}, regarding the geodesic incompleteness of cyclic cosmologies, are not valid}, and that there exist cyclic cosmologies which can have phases of decreasing entropy density, while not violating BGV theorem. \\ \\
Before continuing with the discussion along the announced lines, one should pay attention to another very important critical remark. As very properly pointed out by \cite{prokopec}, the absence of geodesic completeness does not necessary imply the existence of a physical singularity.  This is due to the fact that particles traveling with a non-vanishing velocity on a past incomplete geodesic need to reach super-Plankian energies where the very concept of their trajectory looses it sense (either due to the quantum interaction processes or because they need to collapse to a black hole). This is the reason why such trajectories can not be used for the definition of an initial singularity. As further shown in \cite{prokopec}, when only the particles not reaching the super-Planckian scales are considered, then there will be no singularities if the cosmological scale factor is not vanishing. For this reason, even the cyclic cosmologies which are shown to be geodesically incomplete do not necessarily contain a physical singularity. {Similar considerations and arguments from which several constraints appear can be found in \cite{yoshida, nomura, ling}}.\\ \\
This paper is organized in the following manner: in 2. we review the concept of geodesic incompleteness on FLRW spacetime, review the BGV theorem and geodesic incompleteness of come cyclic models, in 3. we present a class of geodesically complete cyclic models capable of solving the problem of the entropy growth, and we conclude in 4. 

\section{Geodesic incompleteness on FLRW spacetimes}

Here we will give a short review of geodesic incompleteness, specifically in 
Friedmann -
Lemaitre - Robertson - Walker (FLRW) spacetimes. 
The definition of geodesic incompleteness is frequently given as a notion that paths
of observers through spacetime (geodesics) can only be extended for a finite time
as measured by an observer traveling along one, or more technically a property of a
spacetime wherein there exists at least one timelike or null geodesic that cannot be
extended to arbitrarily large values of its affine parameter - such a spacetime 
contains a singularity.
According to the famous singularity
theorems of Hawking and Penrose \cite{ pen1,hawk1,hawk2,hawk3,hawk4, hawkell,hawkpen} 
there is a great indication that there was
a singularity present in the cosmological past, therefore technically we are dealing
with the non-spacelike past geodesic incompleteness.  The formal statement of the
theorem can be concised as follows \cite{wald2}: \textit{Let ($M,g_{\mu \nu}$) be 
a globally hyperbolic spacetime with $R_{\mu \nu} \xi^{\mu} \xi^{\nu}\geq 0$ for all
timelike $\xi^{\mu}$, which will be the case if Einstein's equation is satisfied
with the strong energy condition (SEC) holding for matter. Suppose there exist a 
smooth spacelike Cauchy surface $\Sigma$ for which the trace of the extrinsic 
curvature satisfies $K\leq C< 0$ everywhere, where $C$ is constant. Then all past
directed  timelike geodesics are incomplete.} For the details and proof of the  theorem please consult \cite{wald2}. Hawking and Penrose theorems invoke Einstein's
equation in its statement, and it was in fact shown
that in models of eternal inflation the SEC can be broken and consequently 
circumvent the singularity theorems. However, with the BGV theorem \cite{bgv1} one 
can investigate geodesic completeness from a purely kinematical level. 
For the BGV theorem it will be enough
to adopt a much weaker assumption where the average value of the Hubble parameter,
$H_{av}$, should be grater than zero along some past-directed non-spacelike geodesic.
Let's introduce the FLRW geometry
\begin{equation}
    ds^2=dt^2-a^2(t) (dx^2 + dy^2+dz^2),
    \label{flrwmetrika}
\end{equation}
where $a(t)$ is the scale factor. Firstly, we will inspect the null geodesic kinematics
in this setting. By solving the geodesic equation and using (\ref{flrwmetrika})
one gets $d\lambda \sim a(t) dt$ where $\lambda$ is the affine parameter.
Therefore, if $a(t)$ decreases sufficiently quickly in the past direction,
then $\int a(t) dt$ can be bounded and the maximum of the affine lenght must
be finite \cite{prokopec}. By introducing the Hubble parameter $H(t)=\dot{a}/a$ and 
multiplying it with $d\lambda \sim a(t) dt$ it follows
\begin{equation}
    \int_{\lambda(t_i)}^{\lambda({t_f})}H(\lambda)d\lambda =\int_{a(t_i)}^{a(t_f)}
    \frac{da}{a(t_f)}\leq 1,
    \label{nejednakost}
\end{equation}
where in $d\lambda = [a(t)/a(t_f)] dt$ the parameter $1/a(t_f)$ is used as a 
proportionality constant so that $d\lambda/dt=1$ when $t=t_f$. The inequality
(\ref{nejednakost}) holds if $a(t_i)=0$. By exploiting the definition of
the average quantity of the Hubble parameter we get 
\begin{equation}
    H_{av}=\frac{1}{\lambda(t_f)-\lambda(t_i)}\int_{t_i}^{t_f}H(\lambda)d\lambda
    \leq \frac{1}{\lambda(t_f)-\lambda(t_i)},
    \label{BGV}
\end{equation}
where it is clear that any backward-going null geodesic with $H_{av}>0$ must be
past-incomplete. This is the main result of the BGV theorem which can be generalised
for timelike geodesics, but also for a more general cosmological models, without
assuming homogeneity and isotropy, while assuming a valid definition of the Hubble 
parameter, for details consult \cite{bgv1, bgv2, bgv3, prokopec}. {It could be pointed out that in the original version of the BGV theorem the meaning of the phrase "sufficiently quickly diminishing scale factor" is 
actually not precisely defined. {Furthermore, in the original formulation it is not stressed that the geodesic incompleteness follows from the assumption that $H_{av}(t_i,t_f)$ is bounded from below by $c>0$ when $t_i \rightarrow -\infty$.} The formulation could be made more strict by the slight modification of (\ref{BGV}) in the following manner: {
let $c \in \mathbb{R}$, $c>0$ and $t_f \in \mathbb{R}$, if $\forall t_{i}<t_{f}$, $0<c<H_{av}(t_i,t_{f})$ then it follows $\lambda(t_{f})- \lambda(t_{i})<1/c$, that is, the geodesics are past-incomplete \footnote{{The authors are thankful to the anonymous referee for this proposed generalisation of the BGV theorem}.}.}
\section{Geodesic completeness and cyclic cosmology}}
It was claimed that cyclic cosmologies solving the problem of the entropy growth need to be geodesically incomplete due to the BGV theorem \cite{bgv1, bgv2}. This claim was then recently elaborated and further demonstrated in \cite{bgv3} referring to the model of Ijjas and Steinhardt \cite{steinhardt1}. In order to have geodesic completeness on FLRW spacetime it is necessary that all geodesics are past and future infinite with respect to the proper time, when measured from some arbitrary moment of time. Using the metric of the FLRW spacetime and the geodesic equation it is possible to see \cite{bgv3} that the following integral describing the interval of proper time needs to be divergent:
 \begin{equation}
{\delta \tau= \int^{0}_{-\infty}}\frac{dt}{\sqrt{1+Ca(t)^{-2}}},
\label{integral}
\end{equation}
where $C$ is a constant. Otherwise, the finite value of this integral signals that the geodesic ends at a finite value of the proper time when measured from some arbitrary time taken as "now" -- i.e. that the geodesic is past incomplete.  
If during some cycle of cyclic cosmology we want to have a significant dissipation of entropy density, then we need to have a total increase of the scale factor at the beginning of a new cycle. Thus, the Hubble parameter during expansion, $H_{exp}>0$, and the Hubble parameter during contraction, $H_{contr}<0$, then need to satisfy the following inequality
\begin{equation}
\int_{t_{b}}^{t_{max}} H_{exp} dt \gg - \int_{t_{max}}^{t_{b}+T} H_{contr} dt,
\label{nejednakost2}
\end{equation}
where $t_{b}$ is the time of the bounce at which the cosmological cycle begins,
$t_{max}$ is the time of the maximal scale factor during this cycle, and $T$ is the
period of the considered cosmological cycle. This conditions, of course, means that
the average Hubble rate during this cycle needs to be greater than zero. In the model
of Ijjas and Steinhardt the Hubble parameter is assumed to be strictly periodic,
$H(t)=H(t+T)$, and during each cycle the scale factor is exponentially increased,
$a(t+T)=a(t) e^{N}$. Under such assumptions, the average value of the Hubble parameter
in each cycle will be positive, and so will be the total average Hubble parameter
of the Universe for any number of cycles. Therefore, such cyclic cosmology will be, 
according to BGV theorem, geodesically incomplete. For a more elaborated and 
technical proof starting from these assumptions see \cite{bgv3}. One can also inspect
toy-models for such cyclic cosmology with exponential growth, such as $a(t)=e^{bt}\Big(  
a_0+\sin^2(\omega t) \Big) $, where $b$, $\omega$ and $a_{0}$ are constants, by using such kind
ansatz in integral (\ref{integral}) and showing that the integral is not divergent,
thus demonstrating the geodesic incompleteness. Therefore, one can conclude that the
presented cyclic model is geodesically incomplete.
\\ \\
It would, however, be completely wrong to conclude from the previous discussion that 
cyclic cosmologies in general need to be geodesically incomplete. This is because the
presented type of the model, as well as the proof of its geodesical incompleteness, 
rests on several assumptions which do not need to be satisfied for more general cyclic
cosmologies. Namely:
\begin{itemize}
    \item the Hubble parameter does not need to be periodic in order to have a cyclic (i.e. oscillatory) behavior,
    \item the scale factor does not need to exponentially increase over every cycle,
    \item  the phase of cyclic evolution of the Universe can be a part of more general type of cosmological dynamics,
    \item it is possible that the Universe is essentially empty during some phases of its evolution.
\end{itemize}

\subsection{Entropy growth considerations}
 
The problem with cyclic cosmologies which keep the scale factor periodic over every cycle is basically that this assumption leads to unbounded increase of entropy density. This comes from the fact that the increase of entropy density during contraction leading to a new cycle with periodic $a(t)$ is significantly greater than the possible decrease of entropy density during expansion (while entropy, of course, in general grows both during the contraction and expansion). This is due to the huge increase of entropy during the contracting phase, which is coming from interaction of matter content of the Universe during contraction. {Let us introduce the function of change of entropy density per change of the scale factor
\begin{equation}
    {r_{S}(t)=\frac{1}{\rho_{b}} \frac{d \rho_{s}}{da},}
\end{equation}
 where $\rho_{b}$ is the entropy density at the bounce of the corresponding cycle, $t=t_{b}$.
 Then we can write down the above statement of huge increase in entropy density
 during a periodic cosmological cycle in the following way:
  $\Delta \rho_{S}/ \rho_{b} = \int_{t_{b}}^{tmax} a \cdot r_{S} \cdot H dt + \int_{t_{max}}^{t_{b}+T} a \cdot r_{S} \cdot H dt \gg 1$. Hence, over infinite number of cycles the entropy density diverges. In the remainder of this section we will provide a detailed discussion showing that this is indeed the case. On the other hand, if the scale factor is exponentially increasing and not periodic, and accordingly the inequality (\ref{nejednakost2}) is satisfied, it is possible to have a significant decrease of entropy density over a cycle, {$\Delta \rho_{S}/ \rho_{b} = \int_{t_{b}}^{tmax} a \cdot r_{S} \cdot H_{exp} dt + \int_{t_{max}}^{t_{b}+T} a \cdot r_{S} \cdot H_{contr} dt < 0 $, 
  $|\Delta\rho_S/\rho_b| \gg 1$.} In such a scenario, large regions of the Universe become causally disconnected and characterized by smaller entropy due to exponential expansion, although the total entropy of the Universe continues to grow. But, as we pointed out already, the condition (\ref{nejednakost2}) when applied to the limit $t \rightarrow - \infty$, if $a(t)$ decreases sufficiently quickly in the past direction, leads to $H_{av}>0$ in the past and thus to geodesic incompleteness.} 
 \\ \\
{ In order to demonstrate what was stated above, and to make this discussion of entropy problem more precise and concrete, we should determine cosmological evolution of entropy density, that is make a model for $r_{S}(t)$. This is by no means a simple task, since even the entropy density at present time is open to discussion, not to mention its past and future evolution -- which will also depend on the still unknown details of dark and high energy physics. Apart from that, we also want to stay as general as possible and not limited to specific models. Due to this reasons we can only make some rough estimates based on reasonable physical expectations. The entropy density will in general consist of the contribution from radiation entropy, baryonic and dark matter entropy and dark energy entropy. For the sake of the current discussion we can assume that the dark energy entropy is zero, as it is the case in general relativity. The radiation energy density in thermal equilibrium can be written as \cite{weinberg-entro}
 \begin{equation}
 \rho_{S}^{rad} =\frac{2 \pi^2}{45} g_{*}T^{3},
 \end{equation}
 where $T$ is the temperature.
 As the temperature of the radiation scales like $T \sim 1/a(t)$ it follows that roughly $\rho_{S} \sim 1/a^3(t)$. It can be estimated that the entropy contribution coming from baryons and dark matter is negligible in comparison to radiation entropy today \cite{frampton}, so it can also be safely ignored in the interval from the bounce till today. On the other hand, the matter contribution to entropy will in general rise with respect to the contribution at the bounce time, $t=t_{b}$, due to the second law of thermodynamics, i.e. for the entropy values in the observable Universe we have $S(t)^{mat}=S(t_{b})^{mat} + \Delta S^{mat}(t) $ (with $\Delta S^{mat}(t) >0$) \cite{frampton2}. Thus, we can write $\rho_{S}^{mat}(t) = [1/(V_{0} \times a(t)^3)] [S^{mat}(t_{b})+\Delta S^{mat}(t)]$, where $V_{0}$ is the volume of the observable Universe today (when $a(t=t_{today})=1$). Now we can make an estimate of the average value of change of entropy density per change of the scale factor $\Bar{r}_{S}$ during the expansion from the time of bounce, $t_{b}$, till turnaround, $t=t_{max}$ (while ignoring the effects of all phase transitions and change of effective degrees of freedom during the cosmological evolution -- that is, considering only the dominant scale dependence). We can write the average value of the change of entropy in interval of the expansion of the Universe as:
\begin{equation}
\Bar{r}_{S}^{exp}= \frac{\rho_{S}(t_{max})/\rho_{b}-1}{a(t_{max})-a(t_{b})},  
\end{equation}
 Now, using the above presented dependence on the scale factor, noticing that $a(t_{b}) \ll a(t_{max})$ and ignoring the contribution of matter to entropy density during the bounce, this can be approximated as
 \begin{equation}
\Bar{r}_{S}^{exp} \approx \frac{\Delta S^{mat}(t_{max})}{ V_{0}a(t_{max})^{4} \rho_{b}^{rad}}-\frac{1}{a(t_{max})} ,  
\label{procjena}
\end{equation}
 If there would be no significant increase of the matter contribution to entropy during the expansion, then obviously $\Bar{r}_{S} \approx -1/{a(t_{max})}$. The negative average value of the change of entropy here signifies a decrease of entropy coming from the huge expansion of the Universe from $a(t_{b})$ to $a(t_{max})$. In such a case, obviously $\mid \Bar{r}_{S}\mid \ll 1$ and its value decreases by the increase of the scale factor value at the turnaround. By considering the significant physical parameters, we can indeed see that it is reasonable to assume that the first term in the previous equation is negligible . We can first take into account that $\rho_{b}^{rad}=\rho^{rad}_{S}(today)/ a(t_{b})^{3}$, and also that $\rho^{rad}_{S}(today)\times V_{0}=S^{rad}(today)$. So the first term in equation (\ref{procjena}) can be written as $ [\Delta S^{mat}(t_{max}) \times a(t_{b})^{3}]/[S^{rad}(today) \times a(t_{max})^{4}]$. The value of radiation (and total) entropy in the observable Universe can be estimated to be $10^{88}$ \cite{frampton}, and we can moreover assume that the value of the scale factor during the bounce is associated with the Planck scale, $a(t_{b})=10^{-32}$ and that the Universe will at least continue to expand for some time, say $a(t_{max})=10^5$. Thus we get that the first term in equation (\ref{procjena}) is approximately $\Delta S^{mat}(t_{max}) \times 10^{-204}$. Since the current value of the total entropy of the observable Universe is estimated to be $10^{88}$ today,  it is quite natural to assume that during the expansion of the Universe from today until the turnaround time the growth of entropy contribution coming from matter cannot compensate for such a small number as $10^{-204}$, and this term will be negligible\footnote{{Actually, this estimate is not important for the final estimate of the relative change of the entropy density during the whole cycle, because this term in any case gets cancelled by the corresponding term during the contraction phase.}}. It thus follows, $\Bar{r}_{S}^{exp} \approx -1/{a(t_{max})}$. This value for the average change of entropy density shows that the entropy density has significantly decreased during the expansion from the bounce until the turnaround due to the large increase of the scale factor. 
\\ \\
Now we can determine the average value of the change of entropy density per change of the scale factor during the contraction phase. Let us first consider the simple periodic model of cosmological oscillations, given for instance by $a(t)=a_{0}+\sin^{2}(\omega t)$ or similar periodic functions. We first notice notice that, due to the periodicity of the scale factor, $a(t_{b}+T)=a(t_{b})$, the total entropy density at the new bounce can be written as $\rho_{S}(t_{b}+T)=\rho_{S}^{mat}(t_{b}+T)+ \rho_{S}^{rad}(t_{b}+T)\approx[1/(V_{0} \times a(t_{b})^3)] [S^{mat}(t_{b})+\Delta S^{mat}(t_{b}+T)]+ \rho_{S}^{rad}(t_{b})$, where $\Delta S^{mat}(t_{b}+T)$ is, of course, not a periodic but growing function. 
Using this expressions in the equation for the average value of entropy change per scale factor during contraction
\begin{equation}
\Bar{r}_{S}^{contr}= \frac{\rho_{S}(t_{b}+T)/\rho_{b}-\rho_{S}(t_{max})/\rho_{b}}{a(t_{b})-a(t_{max})},
\end{equation}
and doing the equivalent steps and assumptions as before we finally arrive at:
\begin{equation}
\Bar{r}_{S}^{contr} \approx -\frac{1}{a(t_{max})}- \frac{\Delta S^{mat}(t_{b}+T)}{S^{rad}(today)a_{max}}.
\end{equation}
Taking into account that the relative change of the entropy density per one cosmological cycle is $\Delta \rho_{S}/\rho_{b}= \bar{r}_{S}^{exp} \times (a(t_{max})-a(t_{b}))+ \bar{r}_{S}^{contr} \times(a(t_{b})-a(t_{max}))$, we finally obtain
\begin{equation}
\Delta \rho_{S}/\rho_{b} \approx \frac{\Delta S^{mat}(t_{b}+T)}{S^{rad}(today)} .
\end{equation}
 We see that in the periodic cyclic models there necessary exists a positive change of the entropy density at the end of the cosmological cycle relative to the entropy density at the bounce.  In the model which was used here, this change is approximately equal to the ratio {of}  the increase of matter contribution to entropy during the whole cosmological cycle to the value of radiation entropy today. For the sake of our discussion, it is even not important to make a numerical estimate of this term. It is enough to know that it has a positive value for every cycle, which implies that during the cosmological evolution the entropy density grows with no upper bound.   
 \\ \\
 We now discuss how can this unbounded growth of entropy density over cosmological cycles be avoided in the case of non-periodic oscillatory cosmologies. Let us consider the model of the type $a(t)=e^{\sigma t}(a_{0}+ \sin^{2}(\omega t))$, focusing on our present cosmological cycle. Then, we again have $\Bar{r}_{S}^{exp} \approx -1/{a(t_{max})}$, where $t_{max}$ is the time of the maximal scale factor during the current cycle. Now the total entropy density at the new bounce can be written as $\rho_{S}(t_{b}+T)=\rho_{S}^{mat}(t_{b}+T)+ \rho_{S}^{rad}(t_{b})e^{-3\sigma T}=[e^{-3 \sigma T}/(V_{0} \times a(t_{b})^3)] [S^{mat}(t_{b})+\Delta S^{mat}(t_{b}+T)]+ \rho_{S}^{rad}(t_{b})e^{-3 \sigma T}$. Then, repeating the same calculation, we now have
 \begin{equation}
  \Bar{r}_{S}^{contr} \approx e^{-3 \sigma T}\Bigg[{\frac{1}{e^{\sigma T}a_b-a_{max}}\Big(  \frac{\Delta S^{mat}(t_{b}+T)}{S^{rad}(today)}} + 1\Big) \Bigg],
 \end{equation}
which leads to 
\begin{equation}
\Delta \rho_{S}/\rho_{b} \approx -1 + e^{-3 \sigma T}  \Big[1+\frac{\Delta S^{mat}(t_{b}+T)}{S^{rad}(today)} \Big].  
\label{jeah}
\end{equation}
Generalizing this result to further cycles, where $t=t_{b}+NT$, with $N$ being an integer, we see that the the growth of entropy density starts to diminish with time (since we do not expect that the contribution of the matter to entropy would grow as fast to be an exponential function -- and even in that case, the overall attenuation of the second term in (\ref{jeah}) can be established by the proper choice of $\sigma$ which is a free parameter of the theory). Moreover, at least after a certain time, the entropy density starts to decrease. So, in this type of models there is no divergence, but in fact a decrease of entropy, as we evolve towards the infinite future (which, of course, means that the entropy density was always larger in the past cycles).  Note that, by definition, the asymptotic value for the late times $\Delta \rho_{S}/\rho_{b} \approx -1$ means that the value of entropy density at the end of the cycle is approximately zero (which is, of course, the lower bound for the value of entropy density). 
\\ \\
{We can also generalise the upper result with a more general type of evolving Universe with $a(t+T)=f(T)a(t)$ where $f(T)$ is a monotonically growing function.
In this case, $\rho_S(t)=\rho_S^{mat}(t) +\rho_S^{rad}(t)$ for one whole period  beginning with the bounce becomes $\rho_S(t_b+T)\approx f(T)^{-3}(\Delta S^{mat}(t_b+T)/(V_0 a^3(t_b)) + \rho_b^{rad})$. Then we can calculate  $\Bar{r}_{S}^{contr}$ as
\begin{equation}
    \Bar{r}_{S}^{contr}=\frac{1}{\rho_b}\frac{\rho_S(t_b + T)- \rho_S(t_{max})}{a_bf(T) - a_{max}}
\end{equation}
\begin{equation}
    \Bar{r}_{S}^{contr} \approx f(T)^{-3}\Bigg[{\frac{1}{a_bf(T)-a_{max}}\Big(  \frac{\Delta S^{mat}(t_{b}+T)}{S^{rad}(today)}} + 1\Big) \Bigg],
\end{equation}
and the change of entropy density per $N$ cycles is given as
\begin{equation}
\Delta \rho_{S}(t_b+NT)/\rho_{b} \approx -1 + f(T)^{-3N}  \Big[1+\frac{\Delta S^{mat}(t_{b}+T)}{S^{rad}(today)} \Big]. 
\label{genentropija}
\end{equation}
}From this equation we see see that $f(T)^{-3N}$ needs to decrease sufficiently fast in order to compensate for the matter entropy growth. 
\\ \\
But, while the considered class of toy models, $a(t)=e^{\sigma t}(a_{0}+ \sin^{2}(\omega t))$, solves the problem of unbounded entropy density growth, it is not geodesically complete according to BGV theorem, since it can be checked that for this model $\lim_{t \rightarrow - \infty} H_{av}>0$, and further confirmed by the convergence of the integral (\ref{integral}) for this $a(t)$. Since the positive average value of $H(t)$ during the cosmological cycle is needed in order to have a dominance of expansion over contraction -- which can lead to the attenuation of entropy density, as we have demonstrated in the above model of exponentially growing oscillations -- it may seem that all models solving the entropy growth problem need to be geodesically incomplete.  But we will demonstrate in the next section that this is in fact not the case.  }

\subsection{Geodesically complete cyclic solutions with no entropy problem}
\begin{figure}[H]
    \centering
    \includegraphics[scale=0.73]{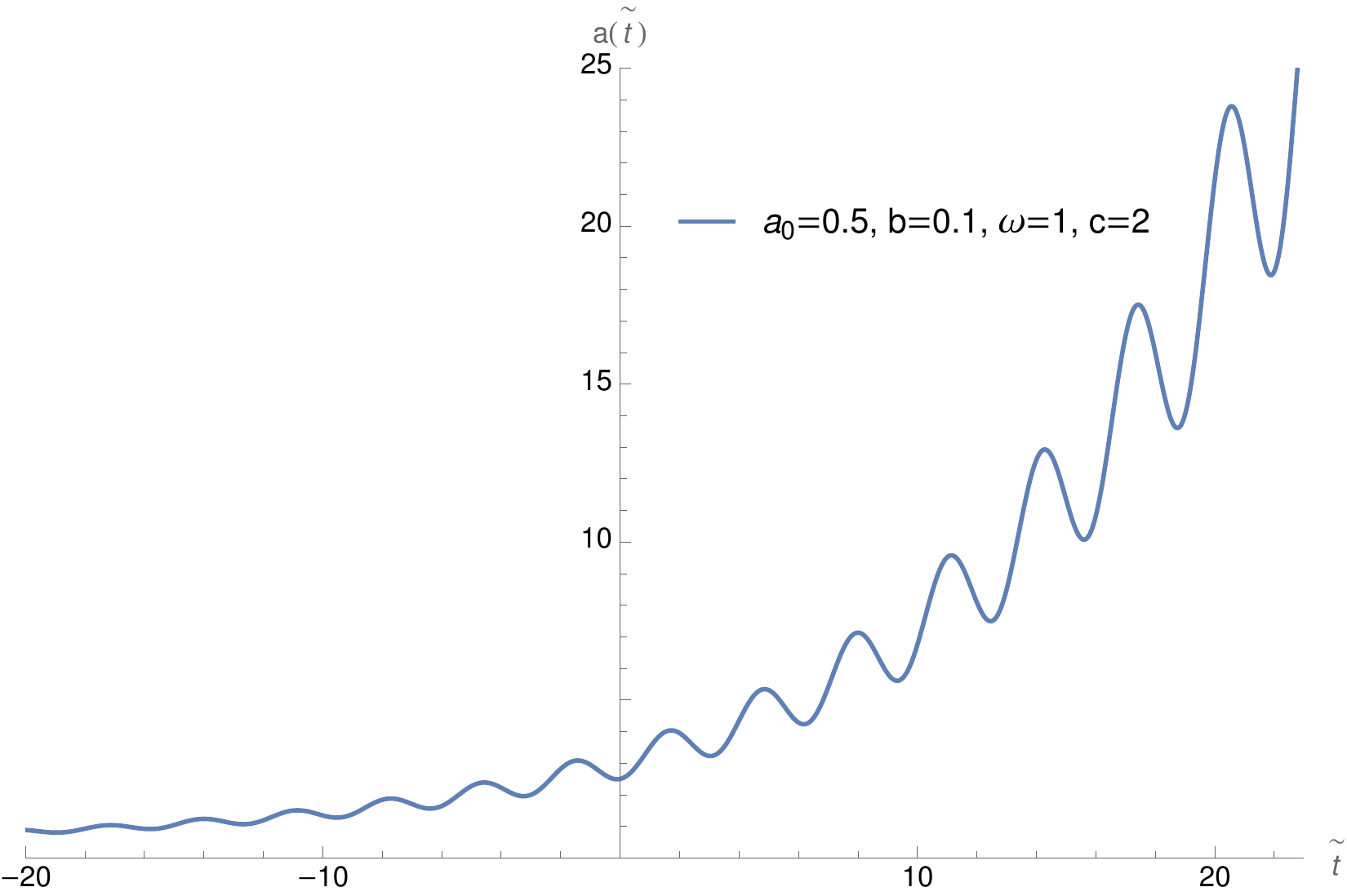}
    \caption{{{Here we present the scale factor for the function (\ref{konst}), where  $a_{0}=0.5$, $\omega=1$, $b=0.1$, $c=2$ and $\tilde{t}$ is time in dimensionless units -- for example $\tilde{t}=H_0t$ with
$H_0$ as the Hubble constant at some moment.}}}
    \label{ljepota}
\end{figure}

\begin{figure}[H]
    \centering
    \includegraphics[scale=0.7]{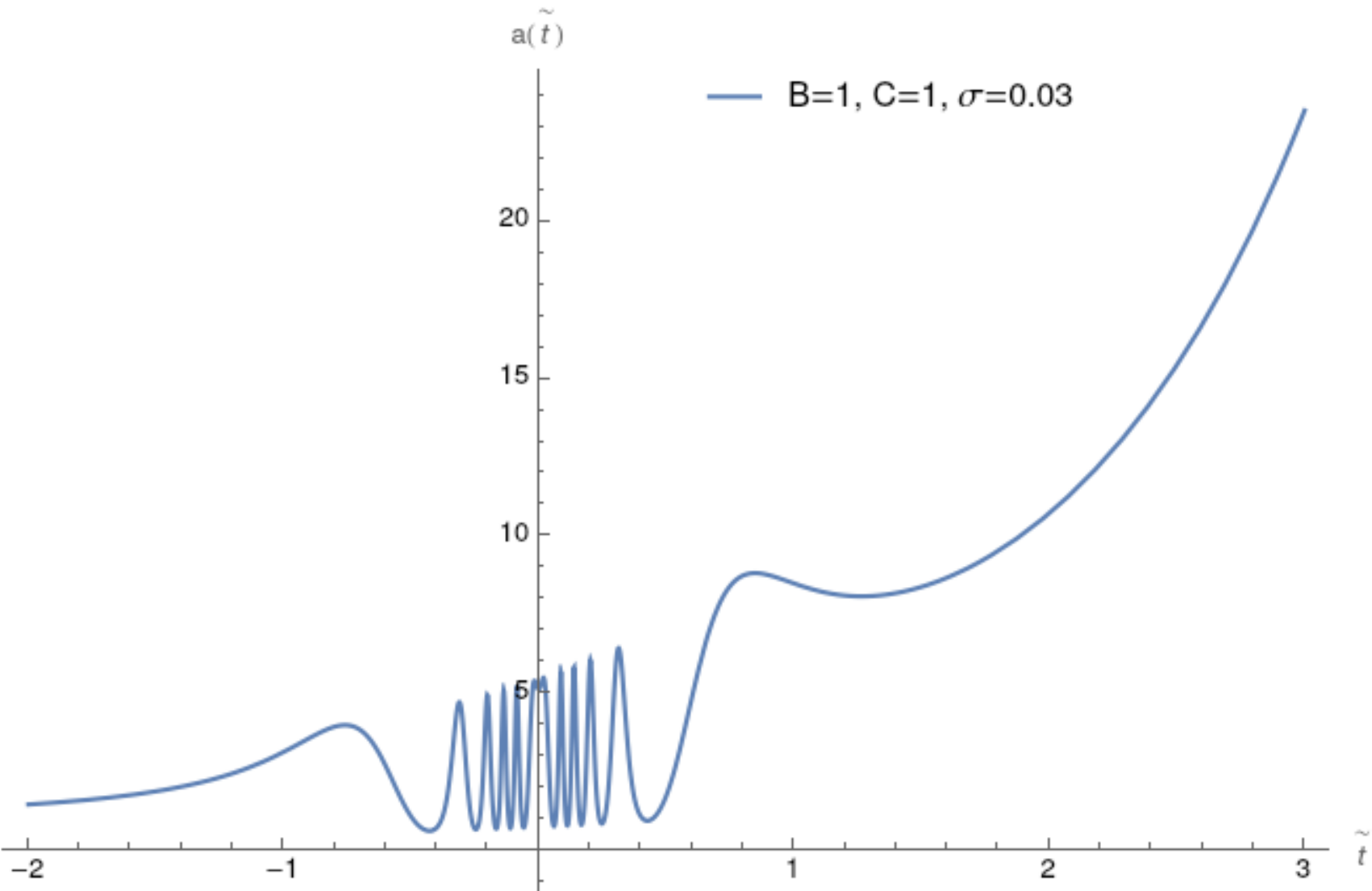}
    \caption{{Here we present the scale factor for the function (\ref{divlja}), where  $A=1$, $B=1$, $c=1$ and $\sigma=0.03$.}}
    \label{zabava}
\end{figure}

{The models which solve the entropy density growth problem and are geodesically complete need to be restricted in their dynamics when $t \rightarrow - \infty$. First of all, if $a(t)$ decreases in the past, then the rate of change which $a(t)$ approaches while $t \rightarrow - \infty$ cannot be greater than zero, since in that case $H_{av}>0$ and the BGV theorem leads to geodesic incompleteness. On the other hand, if $a(t)$ increases while $t \rightarrow - \infty$ there will be an infinite time of diminishing scale factor, and thus increasing entropy density. Namely, as the inequality (\ref{nejednakost2}) is not valid for such cyclic models, while also $a(t) \rightarrow \infty$ for $t \rightarrow - \infty$, this leads to unbounded growth of entropy density in time. This can be clearly seen if we consider a function like $a(t)=e^{ -\sigma t}(a_{0}+ \sin^{2}(\omega t))$ and repeat the final calculation of the previous subsection. We get that the entropy density strongly increases for every cycle in the following manner:
\begin{equation}
\Delta \rho_{S}/\rho_{b} \approx -1 + e^{3 \sigma T}  \Big[1+\frac{\Delta S^{mat}(t_{b}+T)}{S^{rad}(today)} \Big]. 
\end{equation}
Such an effect will be present for every type of evolution manifesting the decrease of the scale factor during time, as can be easily seen by directly generalizing the above expression for any kind of monotonically growing function, $g(t)$ describing the decrease of the scale factor during the subsequent cycles, $a(t_{b}+T)=a(t_{b})/g(T)$, instead of the exponential function. }{In analogy with previous cases,
\begin{equation}
    \Bar{r}_{S}^{contr}=\frac{1}{\rho_b}\frac{\rho_S(t_b + T)- \rho_S(t_{max})}{a_b/g(T) - a_{max}}
\end{equation}
\begin{equation}
    \Bar{r}_{S}^{contr} \approx g(T)^3\Bigg[{\frac{1}{a_b/g(T)-a_{max}}\Big(  \frac{\Delta S^{mat}(t_{b}+T)}{S^{rad}(today)}} + 1\Big) \Bigg],
\end{equation}
and the change of entropy density per $N$ circles for a more general condition
$a(t_{b}+T)=a(t_{b})/g(T)$ where $g(t)$ is a monotonically growing function is given as
\begin{equation}
\Delta \rho_{S}(t_b+NT)/\rho_{b} \approx -1 + g(T)^{3N}  \Big[1+\frac{\Delta S^{mat}(t_{b}+T)}{S^{rad}(today)} \Big]. 
\end{equation}
}
\\ \\
{The remaining possibility of dynamics is that the scale factor decreases while approaching some fixed positive value, $a \rightarrow c$ when $t \rightarrow - \infty$ such that $a(t)>c$, $\forall t < t'$ for some fixed $t' \in  \mathbb{R}$.  In this case it directly follows from definition $H=\dot{a}(t)/a(t)$ that $H_{av} \rightarrow 0$ as $t \rightarrow - \infty$  while also the integral $\int_{-\infty}^{t'}a(t)dt$ diverges (since in this case $\mid \int_{-\infty}^{t'}a(t)dt \mid > c \mid\int_{-\infty}^{t'}dt \mid=\infty$). Thus, in accord with (\ref{BGV}), in this case the affine length diverges, i.e there is no geodesic incompleteness. 
On the other hand, for cyclic models where $a \rightarrow c$ while
$t \rightarrow - \infty$ and which can be modelled as exponentially growing (or, more generally, evolving according to some sufficiently fast growing function) for $t \rightarrow \infty$ 
there is no divergence of entropy density as discussed and shown in equation (\ref{jeah}) and  (\ref{genentropija}).}
\\ \\
All of this can be clearly seen if we consider a simple example of adding a constant to already mentioned oscillatory function with exponential growth. Namely, consider the function 
\begin{equation}
a({t})= a_0+ e^{b{t}}[ \sin^2(\omega {t}) +c],
\label{konst}
\end{equation}
where $b$, $\omega$, $c$ and $a_{0}$ are positive constants. This function is depicted in Fig.(\ref{ljepota})
It is simple to see that $a(t) \rightarrow a_{0}$ when $t \rightarrow -\infty$, while in the same limit $H_{av} \rightarrow 0$. Here the role of the constant $a_{0}$ is of crucial importance -- since, as can be easily checked, if $a_{0}=0$ then $\lim_{t \rightarrow -\infty}H_{av}>0$, while for $a_{0}>0$
$\lim_{t \rightarrow -\infty}H_{av}=0$. 
Also, by adding this function of the scale factor directly to the integral (\ref{integral}) it follows that the integral diverges, again signalling that geodesics on such spacetime are complete. {At the same time, 
as already discussed, the entropy density in such model decreases during the cosmological evolution toward $t \rightarrow \infty$. Note that the presence of constant $a_{0}$ does not play any important role in the entropy density growth considerations since it is enough to consider the relative change $w(t) \equiv a(t)-a_{0}$ as the relevant parameter, and then the discussion basically stays the same as for (\ref{genentropija})  where one can substitute $a(t) \rightarrow w(t)$ and use $f(t)=e^{bt}$, since $w(t+T)=e^{bT}w(t)$}.  
 While the entropy density changes less and less due to the change of the scale factor as we move to $t= -\infty$, there is still a slight contribution to the increase of entropy density due to the spontaneous increase of entropy, assuming the validity of the second law of thermodynamics. In order to  avoid large increase of entropy density, it is sufficient to demand that the decrease of entropy density coming from the slow change of the scale factor needs to compensate for the spontaneous increase of entropy in time. Such change to entropy density due to expansion can be estimated as $\Delta \rho_{S} \approx \rho_{b} r_{S}(a(t)-a_{0})<0$ (as $r_{S}$ is negative since entropy density decreases under dominant effect of expansion). Moreover, as we move to $t \rightarrow \infty$ the entropy density strongly decreases due to the dynamics of exponential expansion. 
\\ \\
In the example considered above, oscillations gradually diminish their amplitude in the past while approaching some constant asymptotic value. Another possible scenario leading to $H_{av} \rightarrow 0$ for $t \rightarrow -\infty$ consists in dynamics of the scale factor where the effective period of oscillations changes during the evolution such that it tends to infinity when $t \rightarrow -\infty$. Consider, for example, the following function of the scale factor
\begin{equation}
a(\tilde{t})= A\Big( 1+Be^{c\tilde{t}}\Big) \exp\Big[{\sin\frac{1}{\tilde{t}^2 + \sigma}\Big]} ,
\label{divlja}
\end{equation}
with positive constants $A$, $B$, $c$ and $\sigma$, which is depicted on Fig.(\ref{zabava}). Here, despite of the difference of dynamics of the scale factor, the limit of $H_{av}$ and $\Delta \rho_{s}$ at $t \rightarrow - \infty$ is the same as in the previous example, so all the discussion regarding the avoiding geodesic incompleteness and entropy divergence is the same as before. From the previous discussions we see that this example consists of tree approximate regimes: slow expansion, oscillations with growing amplitude, and the exponential expansion for $t \rightarrow \infty$, where all this three phases can, for the proper choice of parameters, as discussed in the previous subsection, lead to the decrease of entropy density with time. Although this model does not manifest cyclic evolution over all time, it is still relevant as an example of infinite geodesically complete cosmology with an interval of oscillatory dynamics. Moreover, since it can be easily seen
that in the limit $t \rightarrow \infty$ this scale factor satisfies $a(t) \propto e^{ct}$ this example also demonstrates how inflationary evolution can be made
past geodesically complete.
{\subsection{Physical viability of the proposed models}
It can be asked how realistic are the proposed geodesically complete cyclic cosmologies with no entropy growth problem. In other words, the following question arises: how much of exotic new physics is needed for their support? In addressing this question, we need to first take into account that the main task of this paper was to show that cyclic cosmologies can at the same time be geodesically complete and avoid the entropy growth problem, contrary to some recent claims. For this goal, it was enough to show how to construct a class of counterexamples to the claim that "\textit{BGV theorem makes cyclic cosmologies with decreasing entropy density geodesically incomplete}". Now, when this general claim was proven to be wrong, the physical viability of cyclic cosmologies needs to be determined on case to case basis. However, we can make some general remarks. The cosmologies of the form (\ref{konst}) and similar require the modification of the standard general relativity or can be supported by some appropriate forms of exotic fluids. Even the most simple modified gravity theories, such as $f(R)$ and $f(T)$ gravity have the mathematical structure which is rich enough to support such type of solutions, although the appropriate modification function may not be expressible in analytical form (compare the discussion in \cite{nas1}). Moreover, we can notice that the geodesically complete cosmologies can be constructed from some well known and already existing models avoiding the entropy problem, such as the one proposed in \cite{jen, dva}, if the extra condition, that $a(t) \rightarrow const$ as $t \rightarrow -\infty$ is imposed. 
We thus conclude that geodesically complete cyclic cosmologies in principle do not require any necessary new physics with respect to earlier proposal which were not geodesically complete.} 
    \label{slika3}
    \label{plot1}
    
     %
 
 \section{Conclusion}
 We have started this work by giving a critical examination of relation between concepts of entropy and cyclic cosmologies. We have argued that the arguments related to necessity of entropy growth can not be properly used against the idea of cyclic cosmology, since the status of the entropy concept in general relativity, and its meaning in the evolution of the Universe as a whole, is not at all properly understood. This conclusion is further stressed by the fact that the concept of necessary entropy growth is currently in tension even with the properly tested results of the standard cosmology, namely the CMB spectrum properties.
 \\ \\
 In the further part of the work we have, however, for the sake of the argument, assumed that the concept of entropy growth can be meaningfully applied to cosmology in the usual manner. We then considered the question of geodesic completeness for the cyclic models which in principle can avoid the problems associated with entropy growth. We did this while taking into account the fact that geodesic incompleteness does not necessary imply the existence of physical singularities. 
{After preliminary considerations related to geodesical completeness and entropy growth, we have provided the analysis of the entropy density growth per cycle in the case of periodic cosmologies, as well as in the case of cyclic cosmologies with exponentially growing oscillations, and also for the more general type of growth given by the condition $a(t_{b}+T)=f(T)a(t_{b})$. It was demonstrated that if the scale factor is growing sufficiently fast between the cosmological cycles, the entropy density growth problem can be avoided. We have then furthermore demonstrated that such cosmologies with no entropy density growth problem can be geodesically complete. For this to be the case the following necessary condition $\lim_{t \rightarrow -\infty}a(t)=const.$ needs to be satisfied.} 
 \\ \\
 {We have thus concluded that cyclic cosmologies can be geodesically incomplete for a special set of simple cyclic models, but that it is not possible to use the BGV theorem to conclude that physically realistic cyclic cosmologies are in general geodesically incomplete. Furthermore, we demonstrated that even some very simple models can at the same time be past geodesically complete and solve the problem of entropy growth. Some of the previous discussions in the literature, that made general conclusions on the limitation of cyclic models due to the BGV theorem, are thus not justified. We can conclude that BGV theorem can be very restrictive with respect to geodesic incompleteness in the context of general relativity, where the solutions typically lead to simple dynamics of the scale factor in the form of contraction and expansion, as is the case for inflation. But, for more general gravitational theories (for example, already for simple modifications of GR, such as $f(R)$ and $f(T)$ gravity), which can lead to very rich potential dynamics of the Universe, there will be many models which can be geodesically complete and solve the entropy growth problem.} 

\section*{Acknowledgement}
{Authors are grateful to
two anonymous referees for their very constructive criticism and
discussions, which greatly improved this work.}

\end{document}